\documentclass[sigconf]{acmart}

\usepackage{amsmath}
\usepackage{amsthm}
\usepackage{algorithm}
\usepackage{subcaption}
\usepackage[noend]{algorithmic}

\DeclareMathOperator*{\argmax}{argmax}

\copyrightyear{2026}
\acmYear{2026}
\setcopyright{cc}
\setcctype{by}
\acmConference[RecSys '26]{20th ACM Conference on Recommender Systems}{September 27-October 02, 2026}{Minneapolis, MN, USA}
\acmBooktitle{20th ACM Conference on Recommender Systems (RecSys '26), September 27-October 02, 2026, Minneapolis, MN, USA}
\acmDOI{10.1145/3773078.3831764}
\acmISBN{979-8-4007-2284-4/2026/09}

\begin{document}

\title{MODE: Mutual Optimality in Direct Effects of Reciprocal Recommendations in Matching Markets}

\author{Yoji Tomita}
\email{tomita\_yoji@cyberagent.co.jp}
\orcid{0009-0007-6828-4424}
\affiliation{%
  \institution{CyberAgent, Inc.}
  \city{Tokyo}
  \country{Japan}
}


\begin{abstract}
Matching platforms such as job posting services and online dating platforms have become widely used over the past decade.
For a matching platform to be successful, it is crucial to design appropriate reciprocal recommendation systems (RRSs) that consider the preferences of users on both sides (job candidates and employers) and prevent opportunities from being concentrated too heavily on a few popular users.
However, prioritizing concentration mitigation too much can lead to recommending undesirable results to some individual users, resulting in their dissatisfaction.
In this paper, we formulate the concept of ``optimality of direct effects'' of the recommendation list for an individual user, given the recommendations to other users.
Furthermore, we propose a novel method, MODE, that computes mutually optimal recommendations in direct effects.
Experiments with synthetic and real-world data demonstrate that MODE surpasses other existing methods in terms of mutual optimality of direct effects, exhibits faster processing speeds, and enables a higher expected number of matches.
\end{abstract}

\ccsdesc[500]{Information systems~Recommender systems}
\ccsdesc[300]{Information systems~Social networks}
\ccsdesc[300]{Information systems~Social recommendation}

\keywords{Reciprocal Recommender Systems, Matching Markets, Matching Theory}


\maketitle

\section{Introduction}\label{sec:introduction}

Matching platforms that provide opportunities for users to build new relationships with other users have become widespread in recent decades.
For example, job seekers look for new jobs on job recommendation platforms, and people seek new romantic partners on online dating platforms.
In matching platforms, recommendation systems play a crucial role in the platform's success, and thus \emph{Reciprocal Recommender Systems (RRSs)}, which provide users with recommendation lists of other users, have recently attracted attention from both academia and practitioners.
RRSs must take mutual preferences into account; recommendations should depend not only on the preferences of users receiving recommendation lists but also on the preferences of the users being recommended.
Moreover, they need to avoid concentrations of recommended opportunities and reduce crowding of applications/likes on popular users.
This is because in these platforms there are likely to be small fractions of extremely popular users (e.g., job posts with high wages in job recommendation platforms, or highly attractive users in online dating platforms), but the number of matches that such popular users can have is physically limited.
Therefore, along with estimating users' preferences, the following problem is also important in RRSs: how should we generate ranked recommendation lists by utilizing the estimated preferences of users on both sides?

Several recent works study the above ranking problem of RRSs.
\citet{su2022optimizing} formalize this ranking problem of RRSs, derive a tractable lower-bound function for social welfare (defined as the expected number of matches in the whole platform), and propose optimizing the approximated social welfare via the Frank-Wolfe algorithm~\cite{frank1956algorithm}.
\citet{tomita2023fast} propose constructing recommendation lists based on matching scores derived from a model of matching with transferable utilities in economics.
However, their methods present some practical difficulties for real-world platforms from the perspective of practitioners.
The former method has a scalability issue because it defines the social welfare objective as a function of stochastic recommendation lists and repeatedly solves a linear programming problem with a huge number of variables.
The latter method is not directly based on the RRS framework, and thus it may not work sufficiently well, especially in practical settings.
Therefore, despite these recent contributions, there remains room for improvement in the optimal ranking problem of RRSs.

In this study, we propose a new algorithm called the \emph{Mutually Optimal recommendation in Direct Effects method (MODE)} for the optimal ranking problem in RRSs.
First, within the RRSs framework~\cite{su2022optimizing}, we define two types of effects of recommendation lists on users' utilities: the direct effect, which comes from a user's own recommendation list, and the indirect effect, which comes from recommendation lists for other users.
We argue that direct effects are more important for users' satisfaction with the platform because their own recommendation lists are visible to users, unlike the recommendation lists for others that generate indirect effects.
Second, we define mutually optimal recommendations in direct effects, under which all users' recommendation lists are mutually optimal in direct effects given the recommendation lists of other users.
Third, by utilizing a technique that efficiently computes probability distributions of users' rank in the lists on the other side, we propose the new \emph{MODE} method, which computes (nearly) mutually optimal recommendations  in direct effects.
Finally, we conduct experiments with synthetic data and real-world data from a large online dating platform and show that MODE achieves a higher expected number of matches in all cases, while remaining computationally efficient even on large datasets.

\subsection{Related work}\label{sec:related_work}

\emph{Reciprocal recommender systems (RRSs)}~\cite{pizzato2010recon}, which suggest people to other people, have recently attracted attention from both academia and industry~\cite{palomares2021reciprocal,neve2025reciprocal}.
Main applications of RRSs include online dating~\cite{pizzato2013recommending,xia2015reciprocal,tomita2022matching}, job posting services~\cite{yu2011reciprocal,mine2013reciprocal,almalis2014content}, peer recommendation in education~\cite{potts2018reciprocal}, and social media~\cite{zhang2011intrank}.
Unlike standard item-to-user recommender systems, RRSs need to consider the mutuality of the preferences of users who receive recommendations and users who are recommended.
Therefore, in RRSs, how to utilize estimated preferences of the two sides of users to generate recommendation rankings has been one of the main issues since the early studies, in addition to the estimation of preferences.
An early approach is an aggregation strategy, in which users are recommended in the order of reciprocal scores computed by some operator that aggregates the preferences of users on both sides~\cite{neve2019aggregation,neve2019latent}.
Aggregating operators used in this context are, for example, arithmetic mean~\cite{neve2019latent}, geometric mean~\cite{neve2019latent}, harmonic mean~\cite{pizzato2010recon}, cross-ratio uniform~\cite{appel2017cross,neve2019latent}, and sum of similarities~\cite{almalis2014content}.
Although aggregation approaches are widely used in practice because of their simplicity, their performance depends on the scale of the estimated preferences and the application, and they lack a theoretical foundation.

More sophisticated approaches that apply \emph{matching theory}~\cite{gale1962college,roth1992two,lovasz2009matching} to RRSs have been studied recently.
\citet{saini2019privatejobmatch} explore a multi-match version of the deferred acceptance algorithm, which is a classical matching algorithm~\cite{gale1962college}, for job recommendation.
\citet{su2022optimizing} formalize the ranking recommendation problem in RRSs and propose optimizing an approximated social welfare, defined as the expected number of matches, via a Frank-Wolfe method~\cite{frank1956algorithm}.
\citet{chen2023reducing} and \citet{tomita2023fast} propose to apply the model of \citet{choo2006marries} in the field of matching with transferable utilities~\cite{shapley1971assignment,becker1973theory} to RRSs in online dating.
This paper builds upon \citet{su2022optimizing} and \citet{tomita2023fast} and addresses the remaining challenges discussed in Sections~\ref{sec:introduction} and \ref{sec:recently_proposed_methods}.

Finally, the concept of mutual optimality in direct effects defined in this paper is related to the fairness in RRSs.
Although fairness in recommender systems has attracted growing attention~\cite{li2023fairness}, fairness in RRSs is relatively less studied.
\citet{xia2019we} study fairness in reciprocal recommendation based on Walrasian equilibrium, which is a basic concept in economics.
\citet{do2021two} use the notion of Lorenz efficiency from welfare economics as a criterion of fairness in two-sided matching markets.
\citet{tomita2024fair,tomita2026balancing} consider envy-freeness in the literature on fair division as a metric of fairness in RRSs and propose a fair recommendation method based on Nash social welfare.
Mutual optimality of direct effects defined in this paper is a requirement that no user is sacrificed for others' opportunities, which is inspired by Nash equilibrium~\cite{nash1950equilibrium}.
The relation to Nash equilibrium is discussed in Section~\ref{sec:relation_with_ne}.

\section{Setup}\label{sec:setup}
We first introduce the framework for reciprocal recommendation in matching markets following \citet{su2022optimizing}, and discuss recently proposed methods and the remaining challenges.

\subsection{Framework for ranking recommendation in matching markets}\label{sec:framework}
We consider a job posting service as an example in this paper, but other platforms such as online dating platforms can be considered similarly.
The platform has two types of users.
Let $\mathcal{I} = \{i_1, \dots, i_I\}$ be the set of candidates with size of $I$, and $\mathcal{J} = \{j_1, \dots, j_J\}$ be the set of employers with size of $J$.

\subsubsection*{Actions of Candidates (Proactive Side)}\label{sec:actions_of_candidates}
In the platform, each candidate $i \in \mathcal{I}$ receives a ranked recommendation list of $K (\le J)$ employers $\sigma_i = (\sigma_{i,1}, \dots, \sigma_{i,K})$, where $\sigma_{i,k} \in \mathcal{J}$ is the $k$-th recommended employer to $i$.
Similarly to \cite{su2022optimizing,tomita2023fast}, we assume that each candidate $i$ applies to recommended employers based on \emph{the position-based model (PBM)}~\cite{joachims2017unbiased}.
The candidate $i$ examines the $k$-th recommended employer $\sigma_{i, k}$ with probability $v_{i, k} \in [0, 1]$, and  $i$ applies to $\sigma_{i,k}$ with probability $p_{i, \sigma_{i, k}}$ after examining the profile of $\sigma_{i, k}$.
For $i \in \mathcal{I}$ and $j \in \mathcal{J}$, $p_{i, j} \in [0, 1]$ is $i$'s preference score toward $j$, which was computed by the platform previously via some standard recommendation methods.
The examination probability $v_{i, k}$ is assumed to be non-increasing in $k \in [K]$.\footnote{For $N \in \mathbb{N}$, let $[N] \coloneq \{1, 2, \dots, N\}$.}
The ``inv'' type ($v_{i, k} = 1/k$)~\cite{joachims2017unbiased} or ``log'' type ($v_{i, k} = 1/\log(k + 1)$)~\cite{jarvelin2002cumulated} are typically used in the literature, or examination probabilities can be estimated through the platform's data~\cite{agarwal2019estimating}.
In summary, after the platform chooses the recommendation list $\sigma_i$, the probability $\mathbb{P}_{i,j}^{\sigma}$ that candidate $i$ applies to employer $j$ is\footnote{$\mathbb{I}(A) = 1$ if $A$ is true, $0$ otherwise.}
\begin{equation}\label{eq:P_ij^sigma}
  \mathbb{P}_{i,j}^{\sigma} \coloneq \mathrm{Pr}\left( \text{$i$ applies to $j$} \mid \sigma_i \right) = \sum_{k = 1}^{K} \mathbb{I}\left(\sigma_{i, k} = j\right) \cdot v_{i, k} \cdot p_{i, j}.
\end{equation}

If the platform can randomize the recommendation list for candidates, let $\pi: \mathcal{I} \to \Delta\left(\Sigma_{\mathcal{J}}^{K}\right)$ be a stochastic recommendation policy,\footnote{For a set $A$, $\Delta(A)$ is a set of possible probability distributions over $A$.} where $\Sigma_{\mathcal{J}}^{K}$ is the set of possible recommendation lists of $K$ employers, and $\pi(\cdot \mid i)$ is the probability distribution of the recommendation list for the candidate $i$.
Given a stochastic policy $\pi$, the probability $\mathbb{P}_{i,j}^{\pi}$ that $i$ applies to $j$ is
\begin{align}
  \mathbb{P}_{i,j}^{\pi} &\coloneq \mathrm{Pr}\left(\text{$i$ applies to $j$} \mid \pi\right) \notag\\
  &= \sum_{\sigma_i \in \Sigma_{\mathcal{J}}^{K}} \pi(\sigma_i \mid i) \sum_{k = 1}^{K} \mathbb{I}( \sigma_{i, k} = j ) \cdot v_{i, k} \cdot p_{i, j}\notag\\
  &= \sum_{k = 1}^{K} M^{\pi}_{i, j, k} \cdot v_{i, k} \cdot p_{i, j}, \label{eq:P_ij^pi}
\end{align}
where $M^{\pi}_{i, j, k} \coloneq \mathrm{Pr}_{\sigma_{i} \sim \pi(\cdot \mid i)}\left( \sigma_{i, k} = j \right)$ is the probability that the employer $j$ is recommended to the candidate $i$ in the $k$-th position.

\subsubsection*{Actions of employers (Reactive Side)}\label{sec:actions_of_employers}
After the application phase of candidates, each employer $j$ receives a ranked list $\tau_j$ of candidates who applied to $j$ with the maximum size of $L ~ (\le I)$.
In the ranked list $\tau_j$, top-$L$ applications are selected and ranked in the descending order of the employer $j$'s preference score $q_{j, \cdot}$, which is previously computed.
For simplicity of discussion below, we assume that there is no tie in each employer $j$'s estimated preference, that is, $q_{j, i} \ne q_{j, i'}$ for all $j, i$ and $i' \ne i$.
Similarly to the case of candidates, their actions are based on the PBM.
The employer $j$ examines the application from the $\ell$-th candidate $\tau_{j, \ell} \in \mathcal{I}$ with probability $w_{j,\ell}$, where $w_{j} = (w_{j,1},\dots, w_{j,L})$ is the examination probability vector for the employer $j$.
Conditioning on the examination, $j$ matches with the candidate $\tau_{j, \ell}$ with probability $q_{j, \tau_{j, \ell}}$.
Therefore, the probability $\mathbb{P}_{j,i}^{\sigma}$ with which the candidate $i$ and the employer $j$ are matched conditioning on that $i$ applies to $j$ is:
\begin{align}
  \mathbb{P}_{j,i}^{\sigma} &\coloneq \mathrm{Pr}\left(\text{$i$ matches with $j$}\mid \sigma, ~ \text{$i$ applies to $j$}\right) \notag \\
  &= \sum_{\ell = 1}^{L} \mathrm{Pr}\left(\mathbf{rk}_{j}(i) = \ell \mid \sigma\right) \cdot w_{j, \ell} \cdot q_{j, i}. \label{eq:P_ji^sigma}
\end{align}
where $\mathbf{rk}_{j}(i)$ is the rank of $i$ among candidates applying to $j$ according to $q_{j, \cdot}$.
Given a stochastic policy $\pi$, $\mathbb{P}_{j,i}^{\pi}$ is defined similarly.

\subsubsection*{Social Welfare}\label{sec:social_welfare}
We define the expected matches given recommendation lists $\sigma = (\sigma_i)_{i \in \mathcal{I}}$ as the social welfare function:
\begin{align}\label{eq:sw_sigma}
  \mathrm{SW}(\sigma) \coloneq \sum_{i \in \mathcal{I}}\sum_{j \in \mathcal{J}} \mathbb{P}_{i,j}^{\sigma} \cdot \mathbb{P}_{j,i}^{\sigma}.
\end{align}
Typically, platform's main objective is its maximization: $\max_{\sigma} \mathrm{SW}(\sigma)$.

In the case of a stochastic policy, given a policy $\pi$, the corresponding doubly stochastic matrices $\left(M_{i,j,k}^{\pi}\right)$ and the social welfare $\mathrm{SW}(\pi) \coloneq \sum_{i \in \mathcal{I}} \sum_{j \in \mathcal{J}}\mathbb{P}_{i,j}^{\pi} \cdot \mathbb{P}_{j,i}^{\pi}$ are determined.
Conversely, if $\left(M_{i, j, k}\right) \in [0, 1]^{I\times J \times K}$ such that 1) $\sum_{j \in \mathcal{J}} M_{i, j, k} = 1$ for all $i, k$, and 2) $\sum_{k =1}^{K} M_{i, j, k} \le 1$ for all $i, j$ are given, a corresponding stochastic policy $\pi^{M}:\mathcal{I} \to \Delta\left(\Sigma_{\mathcal{J}}^{K}\right)$ can be computed in a polynomial time via the Birkoff-Neumann decomposition~\cite{birkhoff1946three,singh2018fairness}.
Therefore, we treat $M \in [0,1]^{I\times J \times K}$ satisfying the above conditions just as a stochastic policy below.
Consequently, the optimization problem for a stochastic policy can be written as 
\begin{align}\label{eq:sw_M}
  \max_{M \in [0,1]^{I\times J \times K}}~&\mathrm{SW}(M) \coloneq \sum_{i \in \mathcal{I}} \sum_{j \in \mathcal{J}} \mathbb{P}_{i,j}^{M} \cdot \mathbb{P}_{j,i}^{M}\\
  \mathrm{s.t.} ~~ &\sum_{j \in \mathcal{J}} M_{i,j,k} = 1 ~~ \forall i \in \mathcal{I}, ~ k = 1,\dots,K, \notag \\
  & \sum_{k = 1}^{K} M_{i, j, k} \le 1 ~~ \forall i \in \mathcal{I}, j \in \mathcal{J}. \notag
\end{align}

\subsection{Recently Proposed methods}\label{sec:recently_proposed_methods}

\subsubsection*{Approximated SW optimization~\cite{su2022optimizing}}

Difficulty of direct optimization of $\mathrm{SW}(M)$ in \eqref{eq:sw_M} is due to the terms $\mathrm{Pr}\left(\mathbf{rk}_j(i) = \ell \mid \pi\right)$ in \eqref{eq:P_ji^sigma}, that is the probability distribution of the ranking of $i$ in the employer $j$'s list $\tau_j$.
According to \citet{su2022optimizing}, this probability is
\begin{equation}\label{eq:probability_rank}
  \mathrm{Pr}\left(\mathbf{rk}_{j}(i) = \ell \mid M\right) = \sum_{U \in F_{j,i}^{\ell-1}} \prod_{i' \in U} \mathbb{P}_{i', j}^{M} \prod_{i'' \in A_j(i)\setminus U}\left(1 - \mathbb{P}_{i'', j}^{M}\right)
\end{equation}
where
\begin{equation*}
  A_j(i) = \left\{i' \in \mathcal{I} \mid q_{j, i'} > q_{j, i} \right\}
\end{equation*}
is the set of candidates who are preferred by employer $j$ to $i$, and
\begin{equation*}
  F_{j, i}^{\ell-1} = \left\{U \subset A_{j}(i) \mid \left|U\right| = \ell-1 \right\}
\end{equation*}
is all subsets of $A_j(i)$ with size of $\ell-1$.

Due to the complexity and non-convexity of \eqref{eq:probability_rank}, \citet{su2022optimizing} consider optimizing the approximated social welfare instead of $\mathrm{SW}(M)$.
They derive the lower bound of the social welfare function
\begin{align}
  \underline{\mathrm{SW}}(M) &\coloneq \sum_{i \in \mathcal{I}}\sum_{j\in \mathcal{J}} \mathbb{P}_{i,j}^{M} \cdot w_{j, \ell}\left( \mathbb{E}\left[\mathbf{rk}_{j}(i) \mid M\right] \right)\cdot q_{j,i} \notag \\
  &= \sum_{i \in \mathcal{I}}\sum_{j\in \mathcal{J}} \mathbb{P}_{i,j}^{M} \cdot w_{j, \ell}\left( 1 + \sum_{i' \in A_{j}(i)}\mathbb{P}_{i',j}^{M}\right)\cdot q_{j,i}  \label{eq:approx_sw}
\end{align}
under the assumption of the convexity of examination functions $w_{j, \cdot}$,\footnote{\citet{su2022optimizing} assume that the examination vector $w_{j, \cdot}$ can be extended to the function with the domain of real values $w_j(\cdot)$, and their convexity.} and they propose to optimize $\underline{\mathrm{SW}}(M)$ via Frank-Wolfe method.
The details of their algorithm are described in Section~\ref{sec:details_benchmarks}.

However, their method has several difficulties in order to implement practically in real-world services.
Their method has $I \times J \times K$ variables, and needs to compute gradients and optimize with constraints via linear programming repeatedly.
It imposes a huge computational cost, making it difficult to apply in large real-world platforms.
In addition, their method outputs the doubly stochastic matrices $(M_{i, j, k})$, and thus it is needed to decompose to a stochastic policy $\pi^M$ by Birkhoff--von Neumann algorithm~\cite{birkhoff1946three}, which also requires a large computational cost for each candidate.

\subsubsection*{TU method~\cite{tomita2023fast}}

To address the challenges of \citet{su2022optimizing} discussed above, \citet{tomita2023fast} consider a deterministic policy for the reciprocal recommendation problem.
Based on the literature on matching with transferable utility models~\cite{choo2006marries}, they propose the TU method, which aims to reduce congestion in applications toward popular employers.
Their algorithm is also described in Section~\ref{sec:details_benchmarks}.

Although their method succeeds in some instances in the experimental evaluation, it is not essentially built on the framework of the reciprocal recommendation problem proposed above, and thus it leaves room for improvement.
Especially, the expected number of matches achieved by their method is non-negligibly lower than that of the Approximated SW method in instances with large population biases among employers, which often occur in practical services.

\section{Mutual Optimality in Direct Effects}\label{sec:proposed_method}

In this section, we propose a new deterministic recommendation policy method based on the framework of RRSs.

\subsection{Definition}

Define the candidate $i$'s utility $U_i(\sigma)$ given a policy $\sigma$ as the expected number of matches that $i$ obtain:

\begin{align}
  U_i(\sigma) &= \sum_{j \in \mathcal{J}} \mathbb{P}_{i, j}^{\sigma} \cdot \mathbb{P}_{j, i}^{\sigma} \notag \\
  &= \sum_{j \in \mathcal{J}} \sum_{k = 1}^{K} \underbrace{\mathbb{I}\left(\sigma_{i, k} = j\right)}_{\text{direct effect}} v_{i, k} p_{i, j} \sum_{\ell = 1}^{L} \underbrace{\mathrm{Pr}\left(\mathbf{rk}_{j}(i) = \ell \mid \sigma \right)}_{\text{indirect effect}}  w_{j, \ell}  q_{j, i}. \label{eq:utility}
\end{align}
Utilities given a stochastic policy $U_i(M)$ are defined similarly.
In \eqref{eq:utility}, we see that the effects of a recommendation policy $\sigma$ on $i$'s utility can be divided into two parts.
One is the \emph{direct effect}.
The candidate $i$'s utility depends on the recommendation list he receives.
If employers who $i$ prefers and $i$ is likely to match with are recommended in better ranks, he would get high utility.
The direct effect depends only on his own recommendation list $\sigma_i$, and not on lists for others $\sigma_{-i}$.\footnote{$\sigma_{-i} = (\sigma_{i'})_{i' \in \mathcal{I}\setminus\{i\}}$ is a tuple of recommendation lists excluding the candidate $i$.}
The other is the \emph{indirect effect}.
After $i$ applies to $j$, they are more likely to be matched when $i$ is placed in a better rank in $\tau_j$.
It depends on how many candidates with larger $q_{j, i'}$ apply to $j$, which is affected by recommendation lists for other candidates $\sigma_{-i}$.

Indirect effects are difficult to optimize directly because of their complexity and non-convexity in \eqref{eq:probability_rank}.
On the other hand, given others' recommendation lists $\sigma_{-i}$, direct effects of $i$'s own list $\sigma$ is easy to handle if indirect effects from others' lists can be computed efficiently, whose computation is discussed in a later section.

In addition, optimality of direct effects itself is also important for candidates and platforms in practice.
In general, indirect effects are almost invisible to candidates because they occur through the employers' side and depend on other candidates' lists and actions.
Direct effects from a candidate's own recommendation list are visible to that candidate, and sub-optimality in these effects causes dissatisfaction with the platform because it implies that less attractive or more competitive employers are being recommended to $i$.
In other words, sub-optimality of direct effects means that the candidate is sacrificed for other candidates' opportunities.
Therefore, optimality of direct effects affects candidates' satisfaction toward the platform and the platform's reputation.
In summary, we focus on the optimality of direct effects of recommendation policies.

\begin{definition}{(Optimality of direct effects)}\label{def:optimality_in_direct_effects}
  A recommendation list $\sigma_i^{*}$ is \emph{optimal in direct effects} for a candidate $i$ against others' list $\sigma_{-i}$ if $i$'s list maximizes his own utility given others' list, that is, 
  \begin{equation*}
    \sigma_{i}^{*} \in \argmax_{\sigma_i \in \Sigma^{K}_{\mathcal{J}}} U_i\left(\sigma_i, \sigma_{-i}\right).
  \end{equation*}
\end{definition}
\begin{definition}{(Mutual optimality of direct effects)}\label{def:mutual_optimality}
  A recommendation policy $\sigma^{*}$ is \emph{mutually optimal in direct effects} if for all candidate $i$, $\sigma_{i}^{*}$ is optimal in direct effects against others' list $\sigma_{-i}^*$.
\end{definition}

\subsubsection{Relation with Nash equilibrium in normal form games}\label{sec:relation_with_ne}

Mutual optimality in direct effects has a similarity to (pure strategy) Nash equilibrium (NE)~\cite{nash1950equilibrium} of a non-cooperative normal form game in game theory~\cite{neumann1944theory, nash1951non}.\footnote{A normal form game $G = \left[N, (S_i)_{i \in N}, (u_i)_{i \in N}\right]$ consists of a finite set of players $N = \{1,\dots,n\}$, finite (pure) strategy spaces $S_i$ for each $i \in N$ and utility functions $u_i : S \to \mathbb{R}$. A tuple of pure strategies $s^* = (s_1^*,\dots,s_n^*)$ is Nash equilibrium if $u_i(s_i^*, s_{-i}^*) \ge u_i(s_i, s_{-i}^*)$ for all $i \in N$ and $s_i \in S_i$.}
Indeed, if we interpret candidates $\mathcal{I}$ as a player and possible recommendation lists $\Sigma_{\mathcal{J}}^K$ as a pure strategy space for each candidate, mutual optimality in direct effects corresponds to the pure strategy NE of the reduced game.
If we allow stochastic policies $M$ and extend utilities and mutual optimality in direct effects, it corresponds to mixed strategies and a mixed strategy NE.\footnote{A mixed strategy of the player $i$ is a probability distribution over the pure strategy space $m_i \in \Delta\left(S_i\right)$. A tuple of mixed strategies $m^* = \left(m_1^*,\dots,m_N^*\right)$ is a mixed strategy NE if $U_i(m^*_i, m^*_{-i}) \ge U_i(m_i, m^*_{-i})$ for all $i \in N$ and $m_i \in \Delta\left(S_i\right)$, where $U_i(m) \coloneq \mathbb{E}_{s_1 \sim m_1,\dots, s_N \sim m_N, \mathrm{independently}}\left[u_i(s)\right]$ is $i$'s expected utility.}
Since mixed strategy NE always exist in normal-form games with finite players and strategies~\cite{nash1950equilibrium}, mutual optimal stochastic policies in direct effects also always exist in our framework.
However, computation of mixed strategy NE is hard in general\footnote{It is widely known that computation of mixed strategy NE is PPAD-hard~\cite{daskalakis2009complexity, chen2009settling}.}, and thus it is intractable to solve mutual optimal stochastic policies by general NE solvers such as Lemke-Howson~\cite{lemke1964equilibrium} or Nash Q-learning~\cite{hu2003nash} because candidates' strategy spaces are exponential in the size of employers in our problem.
On the other hand, existence of pure strategy NE is not guaranteed in general.
Therefore, we explore a heuristic approach to derive (nearly) mutually optimal deterministic policy in direct effects, which iteratively computes an optimal list for each candidate.

\subsection{Computation of Ranking Probabilities}

Before we turn to the recommendation algorithm, we need an algorithm to efficiently compute probability distributions of rankings.
Given a policy $\sigma$, we write the ranking probabilities as
\begin{equation*}
  r_{i, j, \ell} \coloneq \mathrm{Pr}\left( \mathbf{rk}_j(i) = \ell \mid \sigma\right)
\end{equation*}
for simplicity.
Then we have the following recursiveness of the ranking probabilities.

\begin{proposition}\label{lem:ranking_probability}
Given a recommendation policy $\sigma$, and let $i_{j, 1}$, $i_{j, 2}$, $\dots$, $i_{j, I}$ be the rearrangement of candidates in the descending order of $j$'s preference $q_{j, \cdot}$ for each employer $j$, that is, $q_{j, i_{j, m}} > q_{j, i_{j, m+1}}$ for all $m \in [I-1]$.
Then, for any employer $j \in \mathcal{J}$, we have
\begin{align*}
  &r_{i_{j, 1}, j, 1} = 1, ~ \mathrm{and} ~ r_{i_{j, 1}, j, \ell} = 0 \hspace{2mm} \mathrm{for} ~ \ell \ge 2,
\end{align*}
for $j$'s most preferable candidate $i_{j, 1}$, and 
\begin{align*}
  &r_{i_{j, m}, j, 1} = \left( 1 - \mathbb{P}_{i_{j, m-1}, j}^{\sigma} \right) r_{i_{j, m-1}, j, 1},\\
  &r_{i_{j, m}, j, \ell} = \left( 1 - \mathbb{P}_{i_{j, m-1}, j}^{\sigma} \right) r_{i_{j, m-1}, j, \ell} +  \mathbb{P}_{i_{j, m-1}, j}^{\sigma}  r_{i_{j, m-1}, j, \ell-1} \hspace{2mm} \mathrm{for} ~ \ell \ge 2
\end{align*}
for other candidates $i_{j, m}$ with $m \ge 2$.
\end{proposition}

The proof is given in Section~\ref{sec:proof_of_lemma}.
This proposition gives us an efficient algorithm to compute the ranking probabilities $(r_{i, j, \ell})$ given a deterministic policy $\sigma$, summarized in Algorithm~\ref{alg:prob_dist_rank}.

\begin{algorithm}[h]
  \caption{RankingProbability}\label{alg:prob_dist_rank}
  \begin{algorithmic}[1]
    \REQUIRE Preferences $p = (p_{i,j}),~ q = (q_{j, i})$, candidates' examination probabilities $v = (v_{i, k})$, and a deterministic policy $\sigma = (\sigma_{i, k})$.
    \ENSURE Ranking probabilities $r = (r_{i, j, \ell} )$.
    \FOR{$j \in \mathcal{J}$}
      \STATE Let $i_{j, 1}, i_{j, 2}, \dots, i_{j, I}$ be the ordering of candidates in descending order of $q_{j, i}$.
      \STATE $r_{i_{j, 1}, j, 1} \leftarrow 1$
      \FOR{$\ell = 2, \dots L$}
        \STATE $r_{i_{j, 1}, j, \ell} \leftarrow 0$
      \ENDFOR
      \FOR{$m = 1, \dots, I-1$}
        \STATE $\mathbb{P}_{i_{j, m}, j}^{\sigma} \leftarrow \sum_{k = 1}^{K} \mathbb{I}(\sigma_{i_{j, m}, k} = j) \cdot v_{i_{j,m}, k} \cdot p_{i_{j, m}, j}$   \label{line:app_prob_in_alg_prob_dist_rank}
        \STATE $r_{i_{j, m+1}, j, 1} \leftarrow (1 - \mathbb{P}^{\sigma}_{i_{j,m}, j}) \cdot r_{i_{j, m}, j, 1}$
        \FOR{$\ell = 2, \dots, L$}
          \STATE $r_{i_{j, m+1}, j, \ell} \leftarrow (1 - \mathbb{P}_{i_{j,m}, j}^{\sigma}) \cdot r_{i_{j, m}, j, \ell} + \mathbb{P}_{i_{j,m}, j}^{\sigma} \cdot r_{i_{j, m}, j, \ell-1}$
        \ENDFOR
      \ENDFOR
    \ENDFOR
    \RETURN $r$
  \end{algorithmic}
\end{algorithm}

For a stochastic policy $M$, we can also compute the ranking probabilities similarly.
Instead of line~\ref{line:app_prob_in_alg_prob_dist_rank} of algorithm~\ref{alg:prob_dist_rank}, the application probability is computed by
\begin{equation*}
  \mathbb{P}_{i_{j, m}, j}^M \leftarrow \sum_{k=1}^K M_{i_{j, m}, j, k} \cdot v_{i_{j, m}, k} \cdot p_{i_{j, m}, j}.
\end{equation*}

\subsection{Proposed method: MODE}

Finally, we explain our proposed method, the \emph{Mutually Optimal recommendation in Direct Effects method (MODE)}.
In our method, we iteratively compute an optimal recommendation list in direct effects for each candidate $\sigma_{i}^t$ against $\sigma_{-i}^{t-1}$.
Given other candidates' list $\sigma_{-i}$, we define the candidate $i$'s gain from each employer $j$ as the match probability between $i$ and $j$ after $i$'s examination:
\begin{align*}
  g_{i, j} \coloneq p_{i, j} \cdot \mathbb{P}_{j, i}^{\sigma_{-i}} = p_{i, j} \sum_{\ell = 1}^{L} r_{i, j, \ell} \cdot w_{j, \ell} \cdot q_{j, i},
\end{align*}
where the ranking probabilities $r_{i, j, \ell}$ are computed by Algorithm~\ref{alg:prob_dist_rank}.
Then, a recommendation list that selects top-$K$ employers with high gains $g_{i, j}$ for $i$ and sorts them in the descending order of the gains is optimal in direct effects for the candidate $i$.
The whole algorithm or our proposed method is summarized in Algorithm~\ref{alg:mode}.

\begin{algorithm}[h]
  \caption{MODE}\label{alg:mode}
  \begin{algorithmic}[1]
    \REQUIRE Preferences $p = (p_{i,j}), ~ q = (q_{j, i})$, examinations $v = (v_{i,k}), ~ w = (w_{j, \ell})$, an initial policy $\sigma^0 = (\sigma^0_{i, k})$, timesteps $T$
    \ENSURE A deterministic policy $\sigma^* = (\sigma^*_{i, k})$
    \FOR{$t = 1, \dots, T$}
      \STATE $r^t \leftarrow$ RankingProbability$(p, q, v, \sigma^{t-1})$    \COMMENT{Algorithm~\ref{alg:prob_dist_rank}}  \label{line:mode_ranking_prob}
      \FOR{$i \in \mathcal{I}$}
        \FOR{$j \in \mathcal{J}$}
          \STATE $g_{i, j}^{t} \leftarrow p_{i, j} \sum_{\ell = 1}^{L} r_{i, j, \ell}^{t} \cdot w_{j, \ell} \cdot q_{j, i}$
        \ENDFOR
        \STATE Let $j_{i, 1}^t, j_{i, 2}^t, \dots, j_{i, J}^t$ be the rearrangement of employers in the descending order of $i$'s gain $g_{i, j}^t$.
        \FOR{$k = 1, \dots, K$}
          \STATE $\sigma_{i, k}^t \leftarrow j_{i, k}^t$
        \ENDFOR
      \ENDFOR
      \IF{$\sigma^{t} = \sigma^{t-1}$}
        \STATE $\sigma^* \leftarrow \sigma^{t}$  \COMMENT{Convergence}  \label{line:mode_converge}
        \STATE \textbf{break}
      \ELSIF{$\sigma^t = \sigma^{s}$ for some $s \le t-2$ {\bf or} $t = T$}
        \STATE $\sigma^* \leftarrow \argmax_{\sigma^{u}:~u=0,\dots,t} SW(\sigma^u)$  \COMMENT{Cycle or MaxStep}  \label{line:mode_cycle}
        \STATE \textbf{break}
      \ENDIF
    \ENDFOR
    \RETURN $\sigma^*$
  \end{algorithmic}
\end{algorithm}

Beginning with an initial policy $\sigma^0$, it repeats to compute the optimal policy in direct effects for each candidate against the previous policies with the maximum iteration step $T$.
If the updates terminate, that is, if $\sigma^{t} = \sigma^{t-1}$, then $\sigma^t$ is mutually optimal in direct effects, and thus the algorithm returns it (line~\ref{line:mode_converge}).
If a policy that appeared earlier in the history appears again, i.e., $\sigma^{t} = \sigma^{s}$ with $s \le t-2$, then the loop has fallen into a cycle.
In this case, as well as when the maximum step $T$ is reached, our algorithm returns the policy that maximizes the expected number of matches within the history (line~\ref{line:mode_cycle}).
As an initial policy, the algorithm can receive a deterministic policy $\sigma^0$ (e.g., randomly shuffled employer lists), or a stochastic policy $M^0$ (e.g., a uniformly random policy $M^0_{i, j, k} = 1/J$ for all $i, j, k$).
The only difference of the latter case in the algorithm is line~\ref{line:mode_ranking_prob} only of the first loop, where it inputs the initial stochastic policy to Algorithm~\ref{alg:prob_dist_rank},
\begin{equation*}
  r^t \leftarrow \mathrm{RankingProbability}(p, q, v, M^0),
\end{equation*}
instead of an initial deterministic policy $\sigma^0$.

\section{Experiments}\label{sec:experiments}
We conduct experiments with synthetic data and with real-world data from an online dating platform to evaluate the MODE method.\footnote{The implementation code for the experiments with synthetic data is available at the following URL: \url{https://github.com/CyberAgentAILab/mode}}

\begin{figure*}[htbp]
  \centering
  \begin{minipage}{\linewidth}
    \centering
    \includegraphics[width=\linewidth]{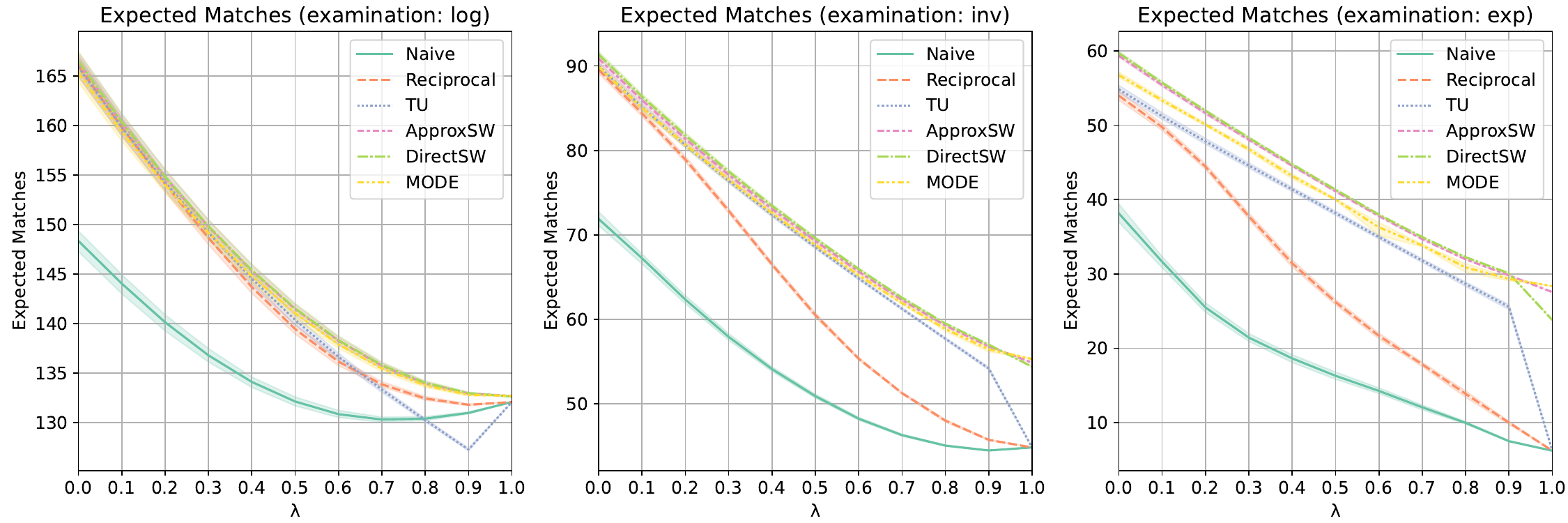}
    \subcaption{The expected number of matches (or social welfare) under each method for various $\lambda$ values and three types of examination vectors.}
    \label{fig:experiment_synthetic_lambda}
  \end{minipage}
  \begin{minipage}{\linewidth}
    \centering
    \vspace{5mm}
    \includegraphics[width=0.666\linewidth]{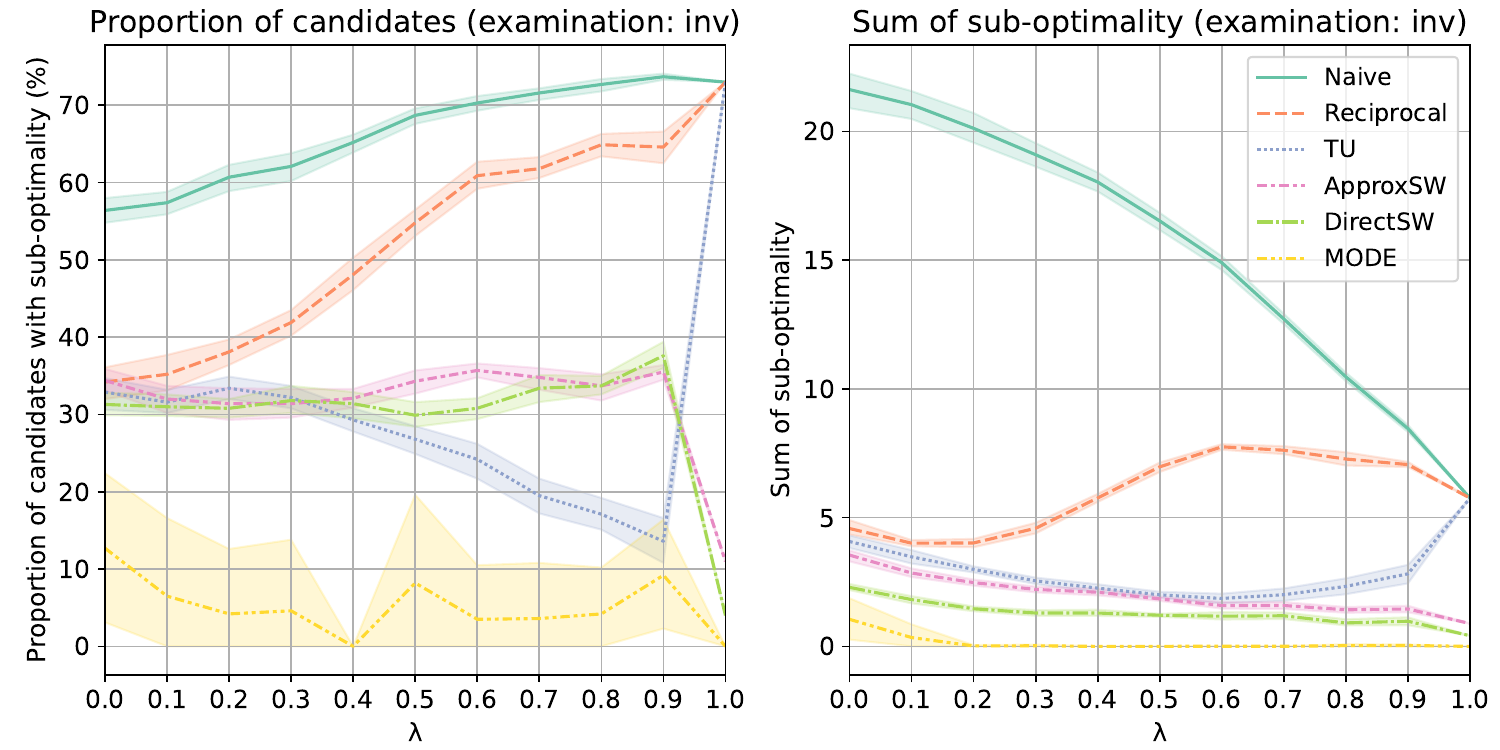}
    \subcaption{Proportion of candidates with non-zero sub-optimality of direct effects, and the sum of the sub-optimality among all candidates.}
    \label{fig:experiment_synthetic_suboptimalty}
  \end{minipage}
  \begin{minipage}{\linewidth}
    \centering
    \vspace{5mm}
    \includegraphics[width=0.666\linewidth]{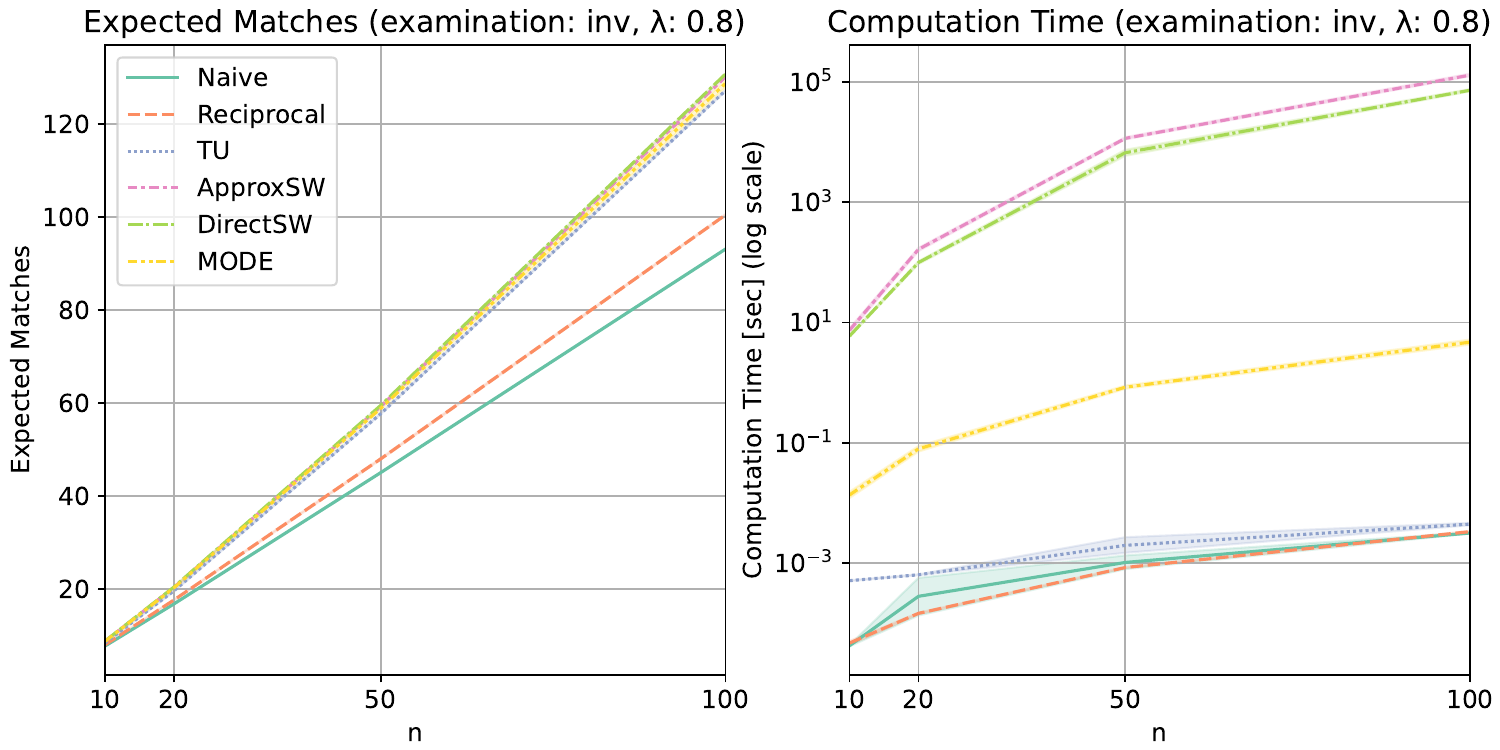}
    \subcaption{The expected number of matches and computation time, varying $n \in \{10, 20, 50, 100\}$.}
    \label{fig:experiment_synthetic_time}
  \end{minipage}
  \caption{Results of experiments with synthetic data. The parameters are $n = 50$, $\lambda = 0.8$, and the examination type is ``inv'' ($v_{i, k} = 1/k$, $w_{j, \ell} = 1/\ell$), unless stated otherwise. Each figure shows the mean and the $90$\% confidence interval over 10 samples.}
  \label{fig:experiment_synthetic}
\end{figure*}

\begin{figure*}[t]
  \centering
  \includegraphics[width=\linewidth]{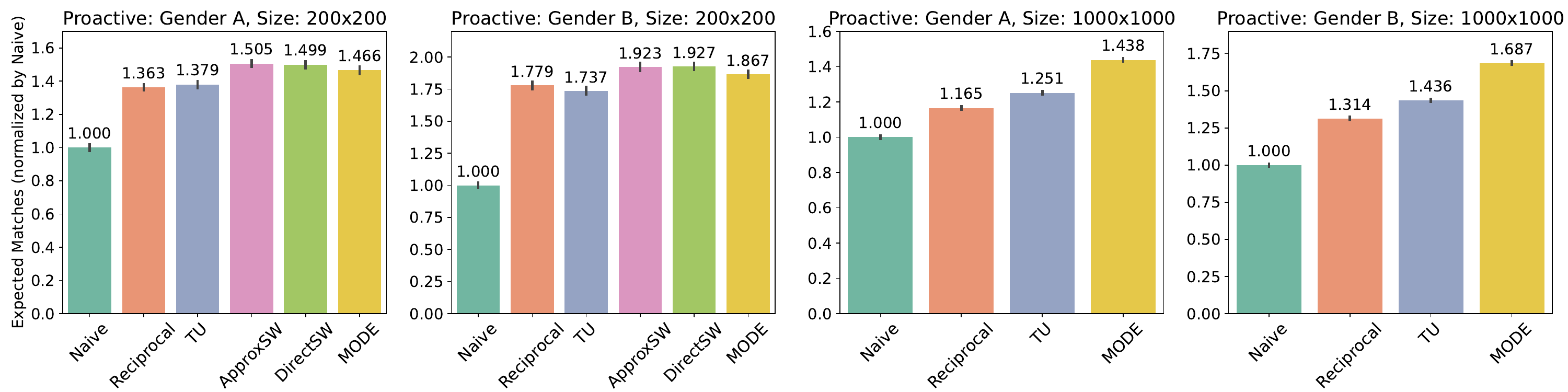}
  \caption{Results of experiments with real-world data. Expected matches are the mean of total matches in $100$ simulations and error bars are 95\% confidence intervals. All results are normalized by the mean value obtained by the Naive method in each case.}
  \label{fig:experiment_real_world}
\end{figure*}

\subsection{Experiments with synthetic data}\label{sec:experiments_with_synthetic_data}

\subsubsection*{Data generation}

In the experiments with synthetic data, we randomly generate a matching market.
The set of employers is $J = n$, and that of candidates is $I = 1.5n$ where $n \in\{10, 20, 50, 100\}$.
Maximum recommendation slots for candidates/employers are set to the number of opposite users, that is, $K = J, ~ L = I$.
Preference scores consist of two parts:
\begin{align*}
  p_{i, j} &= (1-\lambda)\widetilde{p}_{i, j} + \lambda \overline{p}_{j}, ~~ q_{j, i} = (1-\lambda) \widetilde{q}_{j, i} + \lambda \overline{q}_{i}
\end{align*}
for any $i \in \mathcal{I}$ and $j \in \mathcal{J}$.
The first parts are random terms drawn i.i.d. from a uniform distribution, $\widetilde{p}_{i,j}, \widetilde{q}_{j,i} \sim U[0, 1]$.
Second parts are population terms for each candidate/employer, which are defined by $\overline{p}_{j_y} = \frac{J-y}{J-1}$ and $\overline{q}_{i_x} = \frac{I-x}{I-1}$ for $x \in [I]$ and $y \in [J]$.
This means that candidates/employers with small indices are popular among all users on the opposite side.
The parameter $\lambda \in \{0, 0.1, 0.2, \dots, 1.0\}$ controls the level of population congestion.
For examination vectors of candidates and employers, we try three types: ``log'' type $(v_{i, k} = w_{j, k} = 1/\log_2(k+1))$, ``inv'' type $(v_{i, k} = w_{j, k} = 1/k)$ and ``exp'' type $(v_{i, k} = w_{j, k} = \exp(-(k-1)))$.
The default parameters are $n = 50$, $\lambda = 0.8$ and $\text{exam\_type} = \text{``inv''}$ unless stated otherwise.

\subsubsection*{Metrics}

We mainly evaluate recommendation methods, including our proposed method, in terms of the expected number of matches $SW(\sigma)$ defined by \eqref{eq:sw_sigma}, where RankingProbability (Algorithm~\ref{alg:prob_dist_rank}) is used in the computation of $\mathbb{P}_{j, i}^\sigma$ in \eqref{eq:P_ji^sigma}.\footnote{\citet{su2022optimizing} and \citet{tomita2023fast} compute the expected number of matches $SW$ by the Monte Carlo method in their experiments. Since the overall results are not affected by the evaluation method in our experiments, we report the expected number of matches by exact computation as stated above.}

Additionally, we also report the sub-optimality (or regret) of recommendations in direct effects for each candidate, defined by:
\begin{equation}
  \mathrm{SubOptimality}_{i}(\sigma) = \max_{\sigma_i' \in \Sigma_{\mathcal{J}}^{K}} U_i(\sigma'_i, \sigma_{-i}) - U_i(\sigma) \label{eq:suboptimalty}
\end{equation}
for each $i \in \mathcal{I}$ given recommendation lists $\sigma$.
It can also be defined for stochastic recommendations.
If the MODE method terminates with convergence (line~\ref{line:mode_converge} in Algorithm~\ref{alg:mode}), the sub-optimality of direct effects must be zero for all candidates.

For each setting, we generate $10$ samples of matching markets, and report the mean of 10 results for each method.
All experiments are performed simultaneously in 10 cases, on a MacBook Pro 2021 model with Apple M1 Max chip and 32GB memory.

\subsubsection*{Benchmarks}

We compare the MODE method with the following benchmarks:
\begin{itemize}
  \item Naive: For each candidate $i$, it recommends employers in the descending order of $i$'s preference $p_{i, j}$.
  \item Reciprocal: Instead of $i$'s preference alone, it recommends in the descending order of the reciprocal score $rec_{i,j}$ aggregating both sides' preferences. Typical aggregation operator used in literature is simple multiplication: $rec_{i,j} \coloneq p_{i, j} \cdot q_{j, i}$.
  \item TU~\cite{tomita2023fast}: It recommends according to a matching score $\mu_{i, j}$ based on TU matching model.
  \item ApproxSW~\cite{su2022optimizing}: Optimizing the approximated social welfare $\underline{\mathrm{SW}}(M)$ defined in \eqref{eq:approx_sw} by Frank-Wolfe algorithm~\cite{frank1956algorithm}.
  \item DirectSW: Optimizing the social welfare $\mathrm{SW}(M)$ directly via Frank-Wolfe, using RankingProbability by algorithm~\ref{alg:prob_dist_rank}.
\end{itemize}

Details of the TU, ApproxSW, and DirectSW methods are explained in Section~\ref{sec:details_benchmarks}.
Only ApproxSW and DirectSW are stochastic methods, and others including MODE are deterministic methods.

\subsubsection*{Results}

Figure~\ref{fig:experiment_synthetic_lambda} shows the expected number of matches obtained by each recommendation method, varying the crowding-controlling parameter $\lambda$ and the type of examination vector.
In most cases, the expected number of matches achieved by the MODE method is as high as that of the stochastic methods (ApproxSW and DirectSW).
Other deterministic methods (Naive, Reciprocal, and TU) achieve lower expected numbers of matches, especially in large-$\lambda$ cases, which correspond to large population biases that are common in many matching markets such as online dating and job posting services.
On the other hand, the MODE achieves the highest expected matches in such cases.

Figure~\ref{fig:experiment_synthetic_suboptimalty} shows the proportion of candidates with non-zero sub-optimality of direct effects (defined in \eqref{eq:suboptimalty}) and the sum of the sub-optimality of direct effects across all candidates.
Although the MODE method does not always achieve zero sub-optimality of direct effects (which means that MODE does not converge and falls into a cycle in some cases), there are few candidates with sub-optimality (about 10\%) under MODE.
The sum of sub-optimality in direct effects of MODE is almost zero in all cases and substantially lower than that of any other methods.

Figure~\ref{fig:experiment_synthetic_time} shows the performance of each method for various market sizes $n$.
The expected matches of each method increase linearly in $n$, and thus the performances of each method are not affected in market sizes as long as it works.
In terms of computation time, MODE is slightly slower than the other deterministic methods, but substantially faster than the stochastic methods ApproxSW and DirectSW.
At $n = 100$, ApproxSW and DirectSW need more than 10 hours, while MODE needs only a few seconds.
Thus, this experiment suggests the effectiveness of the MODE method on large platforms.

\subsection{Experiments with real-world data}

\subsubsection*{Data and Experimental settings}

In this experiment, we use real-world data from a large online dating platform.
After logging into the platform, users receive recommendation lists of users of the opposite gender in the app.
Users can view the profile details of recommended users and choose to send ``like'' or ``nope''.
After a user sends ``like,'' the recipient can view the sender's profile and choose ``thank'' or ``sorry''.
If ``thank'' is chosen, then they are matched and they can communicate with each other through chats in the app.

For this experiment, we collect behavioral data from about 1,000 users of gender A and 1,000 users of gender B in one of the urban areas where the service is provided.
This dataset contains ``like''/``nope'' behaviors toward recommended users and ``thank''/``sorry'' behaviors toward users who send ``like.''
We estimate proactive-side preferences $(p_{i, j})$ and reactive-side preferences $(q_{j, i})$ from these data by matrix factorization with alternating least squares (ALS)~\cite{koren2009matrix}.

We compute matches under MODE and other benchmarks in 100 simulations, and report the means over the 100 trials.
We show the results only for the ``inv'' examination vectors $(v_{i,k}=w_{j,k}=1/k)$, but the results are not materially affected by different examination types (log or exp).
Since some benchmarks (ApproxSW and DirectSW) are infeasible to compute in the case of 1,000 users for both genders, we also conducted an experiment with data of 200 users for both genders sampled from the original 1,000 users data.
The numbers of recommendation slots ($K$ and $L$) are set to the numbers of opposite gender users.
The experiments are conducted for both the case where gender A users are proactive and the case where gender B users are proactive.

\subsubsection*{Results}

The results are shown in Figure~\ref{fig:experiment_real_world}, which is normalized by the result of Naive in each case, at the platform's request.
In the 200$\times$200 user case, the expected number of matches achieved by MODE is larger than that of the other deterministic benchmarks (Naive, Reciprocal, and TU), and slightly smaller than that of the stochastic benchmarks (ApproxSW and DirectSW), for both the case in which gender A is proactive and the case in which gender B is proactive.
In the 1000$\times$1000 user case, the expected number of matches achieved by MODE is substantially larger (>10\%) than that of any other deterministic method.
These results of experiments with real-world data suggest that the MODE method is effective in platforms especially with a relatively large number of users.

\section{Conclusion}\label{sec:conclusion}

We study the ranking recommendation problem for matching markets.
In matching-market platforms, it is important to design reciprocal recommender systems that mitigate the concentration of recommended opportunities on popular users, which results in congestion and inefficiency.
However, prioritizing congestion mitigation too heavily may cause dissatisfaction among users who receive less attractive recommendations.
We define the optimality of direct effects of recommendation, which means that an individual user receives the best recommendation for their own utility given other users' recommendations and actions.
Using a technique to compute ranking probabilities efficiently, we propose a new method, MODE, which computes a (nearly) mutually optimal recommendation list in direct effects.
Our experiments with synthetic and real-world data show that the MODE method achieves a high expected number of matches in all cases, yields little sub-optimality in direct effects, and runs quickly in comparison with recently proposed methods, suggesting its effectiveness in real-world applications.

\bibliographystyle{ACM-Reference-Format}
\bibliography{references}

\appendix

\begin{figure*}[t]
  \centering
  \begin{minipage}{\linewidth}
      \centering
      \includegraphics[width=\linewidth]{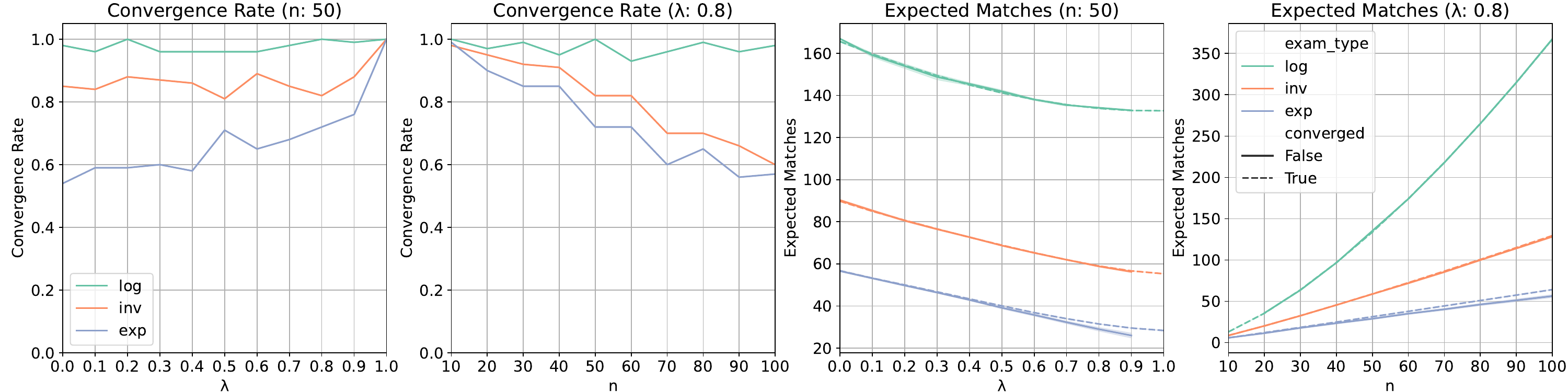}
      \subcaption{Convergence rates of MODE method (2 left figures) and effects of convergence/cycles on social welfare (2 right figures), where expected matches in converged cases are dashed lines and that in cycled cases are solid lines. $100$ experiments are conducted for each parameter setting.}
      \label{fig:additional-convergence}
  \end{minipage}
  \begin{minipage}{\linewidth}
      \centering
      \vspace{2mm}
      \includegraphics[width=\linewidth]{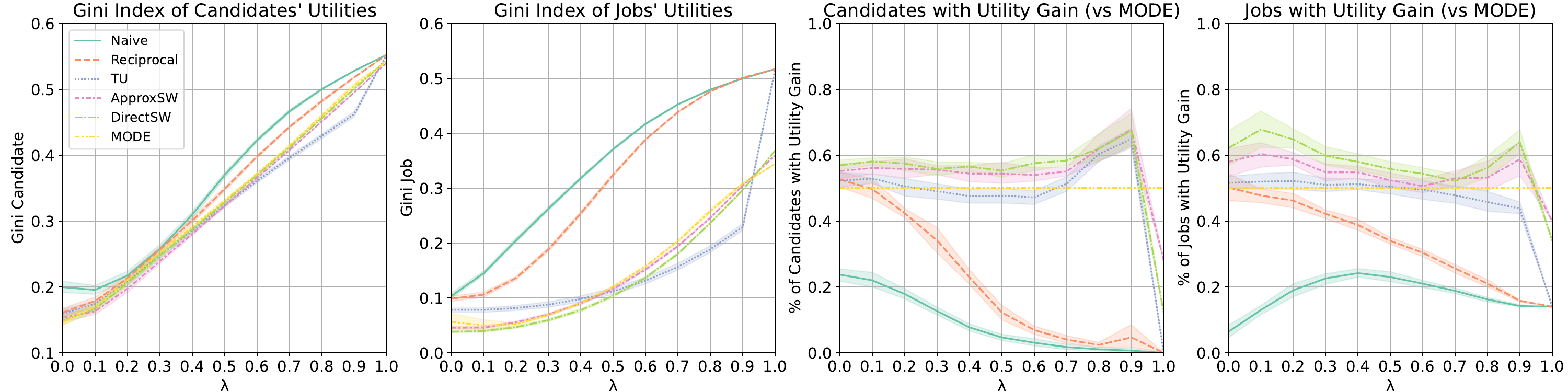}
      \subcaption{Gini index for utilities (expected matches) of candidates and jobs (2 left figures) and percentages of candidates/jobs who would get more expected matches in each method against MODE (2 right figures), where results for MODE are set to $50\%$ for comparison.}
      \label{fig:additional-gini}
  \end{minipage}
  \caption{Results of additional synthetic experiments.}
  \label{fig:additional-results}
\end{figure*}

\section{Proof of Proposition~\ref{lem:ranking_probability}}\label{sec:proof_of_lemma}

Fix an employer $j \in \mathcal{J}$.
A top candidate $i_{j, 1}$ for $j$ must be placed in the 1st rank if $i_{j, 1}$ applies to $j$, because $q_{j, i_{j,1}} > q_{j, i'}$ for any $i' \ne i_{j, 1}$ by definition.
Thus, we have $r_{i_{j,1}, j, 1} = 1$ and $r_{i_{j, 1}, j, \ell} = 0$ for $\ell \ge 2$.

For $m \ge 2$ and $\ell = 1$, from \eqref{eq:probability_rank}, we have
\begin{align*}
  r_{i_{j, m}, j, 1} = \prod_{m' = 1}^{m-1} \left(1 - \mathbb{P}^\sigma_{i_{j, m'}, j}\right)
  = r_{i_{j, m-1}, j, 1} \cdot \left(1 - \mathbb{P}^\sigma_{i_{j, m-1}, j}\right),
\end{align*}
and, for $\ell \ge 2$,
\begin{align*}
  &r_{i_{j, m}, j, \ell} 
  = \sum_{U \in F^{\ell-1}_{j, i_{j, m}}} \prod_{i' \in U} \mathbb{P}_{i', j}^\sigma \prod_{i'' \in A_{j}(i_{j, m})\setminus U} \left(1 - \mathbb{P}_{i'', j}^\sigma\right)\\
  &= ~ \left(\sum_{U \in F^{\ell-1}_{j, i_{j, m-1}}}\prod_{i' \in U} \mathbb{P}_{i', j}^\sigma \prod_{i'' \in A_{j}(i_{j, m-1})\setminus U} \left(1 - \mathbb{P}_{i'', j}^\sigma\right)\right)\cdot \left(1 - \mathbb{P}^\sigma_{i_{j, m-1}, j}\right)\\
    & \hspace{5mm} + \left(\sum_{U \in F^{\ell-2}_{j, i_{j, m-1}}}\prod_{i' \in U} \mathbb{P}_{i', j}^\sigma \prod_{i'' \in A_{j}(i_{j, m-1})\setminus U} \left(1 - \mathbb{P}_{i'', j}^\sigma\right)\right)\cdot \mathbb{P}^\sigma_{i_{j, m-1}, j}\\
  &= r_{i_{j, m-1}, j, \ell} \left(1 - \mathbb{P}^\sigma_{i_{j, m-1}, j}\right) + r_{i_{j, m-1}, j, \ell-1} \mathbb{P}^\sigma_{i_{j, m-1}, j}.
\end{align*}
\qed

\section{Details of TU, ApproxSW, DirectSW}\label{sec:details_benchmarks}

\subsubsection*{TU method~\cite{tomita2023fast}}
TU method is a deterministic method proposed which recommends employers to candidates in the descending order of matching scores $\mu_{i, j}$ computed by Algorithm~\ref{alg:tu}, based on the matching with transferable utility model.
In our experiments, we set $\beta = 1.0$ and $T = 1000$, and we terminate the iteration if the updates of $a_i, b_j$ are smaller than $\epsilon=$ 1e--9, while the algorithm terminates in fewer than 1000 iterations in all experiments.

\begin{algorithm}[htbp]
  \caption{TU Matching based recommendation~\cite{tomita2023fast}}\label{alg:tu}
  \begin{algorithmic}[1]
    \REQUIRE Preferences $p = (p_{i,j}), ~ q = (q_{j, i})$, a scale parameter $\beta ~ (> 0)$, timesteps $T$
    \ENSURE A matching score $\mu = (\mu_{i, j})$
    \STATE Initialize $a_i \leftarrow 1, ~ b_j \leftarrow 1$ for all $i \in \mathcal{I}, ~ j \in \mathcal{J}$
    \FOR{$t = 1, \dots, T$}
      \FOR{$i \in \mathcal{I}$}
        \STATE $x_i \leftarrow \frac{1}{2} \sum_{j \in \mathcal{J}} \exp\left( \frac{p_{i,j} + q_{j, i}}{2\beta} \right) b_j$
        \STATE $a_i \leftarrow \sqrt{1 + x_i^2} - x_i$ 
      \ENDFOR
      \FOR{$j \in \mathcal{J}$}
        \STATE $y_j \leftarrow \frac{1}{2}\sum_{i \in \mathcal{I}} \exp\left(\frac{p_{i,j} + q_{j, i}}{2\beta}\right) a_i$
        \STATE $b_j \leftarrow \sqrt{1 + y_j^2} - y_j$
      \ENDFOR
    \ENDFOR
    \STATE $\mu_{i, j} \leftarrow \exp\left(\frac{p_{i,j} + q_{j, i}}{2\beta}\right) a_i b_j$ for all $i \in \mathcal{I}, ~ j \in \mathcal{J}$.
    \RETURN $\mu$
  \end{algorithmic}
\end{algorithm}

\subsubsection*{ApproxSW method~\cite{su2022optimizing}}

ApproxSW method optimizes the approximated social welfare $\underline{\mathrm{SW}}(M)$ (defined in \eqref{eq:approx_sw}) via Frank-Wolfe algorithm~\cite{frank1956algorithm}.
The algorithm is summarized in Algorithm~\ref{alg:approx_sw}.
In our experiments, we set $\eta_t = 0.2, T = 100$, and we terminate the iterations if the updates of $\underline{\mathrm{SW}}$ are smaller than 1e--3.
We use CVXPY~\cite{diamond2016cvxpy} with the SCS solver~\cite{o2016conic}.

\begin{algorithm}[htbp]
  \caption{Approximated SW optimization via Frank-Wolfe~\cite{su2022optimizing}}\label{alg:approx_sw}
  \begin{algorithmic}[1]
    \REQUIRE Preferences $p = (p_{i, j}), ~ q = (q_{j, i})$, examination probabilities $v = (v_{i, k}), ~ w = (w_{j, \ell})$, timesteps $T$, learning rates $\eta = (\eta_t)$
    \ENSURE A stochastic recommendation policy $M = (M_{i, j, k})$
    \STATE Initialize $M_{i, j, k} = \frac{1}{J}$ for all $i \in \mathcal{I}, j \in \mathcal{J}, k \in [K]$
    \FOR{$t = 1, \dots, T$}
      \STATE $X^* \in \argmax_{X_{i,j,k} \in [0,1]}~\sum_{i, j, k} \frac{\partial \underline{\mathrm{SW}}(X)}{\partial X_{i,j,k}} X_{i, j, k}$ \hspace{1mm} \COMMENT{$\underline{\mathrm{SW}}$ is \eqref{eq:approx_sw}} \\ \hspace{9mm} s.t. $\sum_{j'}X_{i,j',k} = 1 ~\mathrm{and}~ \sum_{k' = 1}^{K} X_{i,j,k'} \le 1$ for all $i, j, k$. \label{line:alg_approx_sw_differentiation}
      \STATE $M \leftarrow (1 - \eta_t)M + \eta_t X^*$
    \ENDFOR
    \RETURN $M$
  \end{algorithmic}
\end{algorithm}

\subsubsection*{DirectSW method}

DirectSW is a stochastic method that directly optimizes the social welfare $\mathrm{SW}(M)$ via Frank-Wolfe algorithm.
It is similar to ApproxSW, but it uses the exact $\mathrm{SW}(M)$ in line~\ref{line:alg_approx_sw_differentiation} of Algorithm~\ref{alg:approx_sw} instead, which is computed by Algorithm~\ref{alg:prob_dist_rank}.
The derivative is obtained by automatic differentiation in PyTorch in our implementation.
It has no theoretical guarantee of convergence to the global optimum since $\mathrm{SW}(M)$ is not concave, but it achieves a relatively high expected number of matches in our experiments.
Similarly to ApproxSW, we set $\eta_t = 0.2, T = 100$, termination criteria 1e--3, and we use CVXPY with the SCS solver for optimization.

\section{Additional Experimental Results}


\subsection{Convergence/Cycles in MODE}
The MODE method (algorithm~\ref{alg:mode}) returns a mutually optimal policy in direct effects if it converged (line~\ref{line:mode_converge}).
Otherwise it falls into cycle, it returns a policy maximizing $SW$ in history.
To study convergence rates and its effects on social welfare, we conducted synthetic experiments with $100$ samples for each parameter setting, which is shown in Figure~\ref{fig:additional-convergence}.
The MODE method converges in $82$\% of samples in default setting ("inv", $n=50$, $\lambda=0.8$), but converge rates are affected by the type of examination and market size $n$.
However, convergence or cycles have little effects on social welfare.

\subsection{Initial policy}

To examine the effects of initial policy in the MODE method, we try a uniformly random stochastic policy, a randomly chosen deterministic policy, and other basic deterministic methods as the initial policy.
Table~\ref{tab:initial-policy} presents the results in the default parameter setting, and it shows that the initial policy does not materially affect performance.

\subsection{Other fairness statistics}

\begin{table}[t]
  \centering
  \caption{Results of MODE with various initial policies. Means and standard deviations by 10 trials are reported.}
  \label{tab:initial-policy}
  \begin{tabular}{lrr}
  \toprule
  Initial Policy & Expected Matches & Sum sub-optimality \\ \midrule
  Uniform & $58.8681 ~(\pm 0.4942)$ & $0.0539 ~(\pm 0.0943)$ \\
  Random  & $58.8484 ~(\pm 0.5078)$ & $0.0721 ~(\pm 0.1319)$ \\
  Naive   & $58.8868 ~(\pm 0.1650)$ & $0.0748 ~(\pm 0.1696)$ \\
  Reciprocal & $58.9132 ~(\pm 0.4511)$ & $0.0466 ~(\pm 0.0923)$ \\
  TU      & $58.8965 ~(\pm 0.4467)$ & $0.1470 ~(\pm 0.4836)$ \\
  \bottomrule
  \end{tabular}
\end{table}

Finally, we report two other fairness metrics in our synthetic experiments.
Gini index~\cite{gini1936measure} is a widely used metric in economics which measures fairness of income distributions, where low Gini index means fair distribution.
Following \citet{tomita2026balancing}, we use the Gini index of users' utilities (expected matches) as other fairness metrics.
In addition, for each benchmark we report fractions of users who would get more expected matches than the MODE method.
Results are shown in Figure~\ref{fig:additional-gini}.
MODE is superior to Naive and Reciprocal method in both metrics, and comparison to other benchmarks depend on a parameter setting.

\end{document}